\documentclass[aps,prl,10pt,twocolumn,nofootinbib,groupedaddress,superscriptaddress,floatfix]{revtex4-1}

\usepackage[colorlinks=true,urlcolor=blue,linkcolor=,citecolor=blue]{hyperref}
\usepackage{amsmath,amssymb,amstext,amsbsy,amsfonts,amsthm,graphicx,microtype,booktabs}
\usepackage[capitalize]{cleveref}

\newcommand{\ud}{\textrm{d}}
\newcommand{\Mpl}{M_\mathrm{pl}}

\begin{document}

\title{Constraining axion inflation with gravitational waves across 29 decades in frequency}

\author{Peter Adshead}
\email{adshead@illinois.edu}
\affiliation{Department of Physics, University of Illinois at Urbana-Champaign, Urbana, Illinois 61801, USA}

\author{John T. Giblin, Jr.}
\email{giblinj@kenyon.edu}
\affiliation{Department of Physics, Kenyon College, Gambier, Ohio 43022, USA}
\affiliation{CERCA/ISO, Department of Physics, Case Western Reserve University, Cleveland, Ohio 44106, USA}

\author{Mauro Pieroni}
\email{mauro.pieroni@uam.es}
\affiliation{Instituto de F\'{\i}sica Te\'orica UAM/CSIC, Calle Nicol\'as Cabrera 13-15, Cantoblanco E-28049 Madrid, Spain}
\affiliation{Departamento de F\'{\i}sica Te\'orica, Universidad Aut\'onoma de Madrid (UAM) Campus de Cantoblanco, 28049 Madrid, Spain}
\affiliation{Theoretical Physics, Blackett Laboratory, Imperial College, London, SW7 2AZ, United Kingdom}

\author{Zachary J. Weiner}
\email{zweiner2@illinois.edu}
\affiliation{Department of Physics, University of Illinois at Urbana-Champaign, Urbana, Illinois 61801, USA}

\begin{abstract}
    We demonstrate that gravitational waves generated by efficient gauge preheating after axion inflation generically contribute significantly to the effective number of relativistic degrees of freedom $N_\mathrm{eff}$.
    We show that, with existing Planck limits, gravitational waves from preheating already place the strongest constraints on the inflaton's possible axial coupling to Abelian gauge fields.
    We demonstrate that gauge preheating can completely reheat the Universe regardless of the inflationary potential.
    Further, we quantify the variation of the efficiency of gravitational wave production from model to model and show that it is correlated with the tensor-to-scalar ratio.
    In particular, when combined with constraints on models whose tensor-to-scalar ratios would be detected by next-generation cosmic microwave background experiments, $r\gtrsim 10^{-3}$, constraints from $N_\mathrm{eff}$ will probe or rule out the entire coupling regime for which gauge preheating is efficient.
\end{abstract}

\maketitle

Reheating is a critical component of a complete, fundamental theory of inflation~\cite{Guth:1980zm, Linde:1981mu, Albrecht:1982wi, Linde:1983gd, Akrami:2018odb}.
Though cosmic microwave background (CMB) observations have yet to determine a unique model of inflation, there must be a mechanism which couples the inflationary sector to the standard model (whether directly or via other relativistic species) to transition the Universe from the cold state left by inflation to the hot Big Bang~\cite{Traschen:1990sw, Shtanov:1994ce, Kofman:1994rk, Kofman:1997yn}.
In the standard or elementary reheating scenario, perturbative decays deplete the homogeneous inflaton condensate into relativistic degrees of freedom which thermalize in time for Big Bang Nucleosynthesis.
Many coupling structures also exhibit a regime of \textit{preheating}, an initial stage of reheating characterized by the exponential production of particles via nonlinear effects (see~\cite{Amin:2014eta,Allahverdi:2010xz} for reviews).

The rapid production of inhomogeneities during preheating typically sources a significant gravitational wave background~\cite{Khlebnikov:1997di, Easther:2006gt, Easther:2006vd, Easther:2007vj, GarciaBellido:2007dg, Dufaux:2007pt, Dufaux:2010cf, Bethke:2013aba, Figueroa:2013vif, Bethke:2013vca, Figueroa:2016ojl, Figueroa:2017vfa}.
On the one hand, unless the inflationary scale is especially low, this stochastic gravitational wave background would reside at high frequencies (typically $10^{6} \lesssim f \lesssim 10^9$) which are far out of reach of present~\cite{TheLIGOScientific:2014jea,TheVirgo:2014hva} and planned~\cite{Somiya:2011np,Audley:2017drz,Punturo:2010zz} direct-detection experiments.
On the other hand, subhorizon gravitational waves gravitate as radiation, allowing their contribution to the effective number of neutrino species $N_\mathrm{eff}$ to be constrained by CMB experiments~\cite{Maggiore:1999vm}.
Indeed, Planck already limits the net energy density in gravitational waves (i.e., all relativistic degrees of freedom beyond the standard model) to $\Omega_{\mathrm{gw},0} h^2 \lesssim 1.2 \times 10^{-6}$~\cite{Pagano:2015hma}.
Next-generation experiments, such as CMB-S4~\cite{Abazajian:2019eic}, will limit $\Omega_{\mathrm{gw},0} h^2 \lesssim 1.68 - 3.36 \times 10^{-7}$, while combined forecasts even project $\Omega_{\mathrm{gw},0} h^2 \lesssim 7.6 \times 10^{-8}$ at $2 \sigma$~\cite{Pagano:2015hma}.

In this Letter and its companion article~\cite{Adshead:2019lbr} we demonstrate that the gravitational waves produced during preheating lead to stringent constraints on the coupling between a pseudoscalar inflaton and gauge fields~\cite{Turner:1987bw, Garretson:1992vt, Anber:2006xt}.
While it has been recently demonstrated that preheating~\cite{Adshead:2015pva,Adshead:2016iae} leads to a potentially important gravitational wave background~\cite{Adshead:2018doq} in these models, in this work we demonstrate that such significant gravitational wave production is \textit{generic} to these models, and we explore the dependence of preheating and the associated gravitational wave production on the details of the potential.
We establish that, regardless of the model of inflation, regimes which efficiently reheat the Universe through preheating alone necessarily result in a detectable level of gravitational waves through their contribution to $N_\mathrm{eff}$.
Varying the scale and shape of the potential alters the efficiency of gravitational wave production, and models with larger tensor-to-scalar ratios exhibit the most efficient gravitational wave production from preheating.
In particular, for models whose tensor-to-scalar ratio would be detected by CMB-S4, $r\gtrsim 10^{-3}$, we show that the projected improvement on the $N_\mathrm{eff}$ constraints would rule out the entire regime for which preheating is $\gtrsim 80\%$ efficient.
In fact, for these models, we show that Planck~\cite{Ade:2015lrj,Akrami:2018odb} already places stringent (model-dependent) bounds on the axion--gauge-field coupling strength.

\textit{Background and models.}---Axions are a particularly appealing candidate as inflaton fields, as their (approximate) shift symmetry protects the flatness of the potential required for slow-roll inflation.
This shift symmetry also severely limits the possible couplings of the inflaton to other sectors.
We couple the axion to the Chern-Simons density of a U(1) gauge field, described by the action
\begin{align}
    \begin{split}
    \label{eqn:action}
        S
        &= \int \ud^4 x \sqrt{-g} \Bigg[
            \frac{\Mpl^2}{2} R - \frac{1}{2} \partial_\mu \phi \partial^\mu \phi - V(\phi) \\
        &\hphantom{={} \int \ud^4 x \sqrt{-g} \Bigg[ }
            - \frac{1}{4} F_{\mu\nu} F^{\mu\nu} - \frac{\alpha}{4 f} \phi F_{\mu\nu}\tilde{F}^{\mu\nu}
            \Bigg].
    \end{split}
\end{align}
Here $\phi$ is the pseudoscalar inflaton (axion), $A_\mu$ is a U(1) gauge field with field strength $F_{\mu\nu} \equiv \partial_\mu A_{\nu} - \partial_\nu A_{\mu }$ whose dual is $\tilde{F}^{\mu \nu} \equiv \epsilon^{\mu \nu \alpha \beta} F_{\alpha \beta} / 2$, and we denote by $\Mpl = 1/\sqrt{8 \pi G_N} = 2.44 \times 10^{18} \, \mathrm{GeV}$ the reduced Planck mass.
The axion--gauge-field coupling is parametrized by $\alpha / f$.

We work with the mostly plus, conformal Friedmann-Lema\^{\i}tre-Robertson-Walker (FLRW) metric, for which the conformal Hubble parameter is $\mathcal{H} \equiv \partial_0 a / a$.
The dynamics of this system are given by the equations of motion for the gauge field and axion,
\begin{align}
    \label{eq:A_i-eom}
    \partial_0^2 A_i - \partial_j \partial_j A_i - \frac{\alpha}{f} \partial_\alpha \phi \tilde{F}^{i\alpha}
    &= 0, \\
    \label{eq:phi-eom}
	\partial_0^2 \phi - \partial_i \partial_i \phi + 2 \mathcal{H} \partial_0 \phi + a^2 \frac{\ud V}{\ud \phi}
    &= - a^2 \frac{\alpha}{4 f} F_{\mu \nu} \tilde{F}^{\mu \nu},
\end{align}
together with the Friedmann equations for the background metric, which are solved self-consistently.

At the level of the homogeneous background, the axion--gauge-field interaction induces tachyonic production of (polarized) gauge bosons during inflation, which results in rich phenomenology, including non-Gaussianities~\cite{Barnaby:2010vf, Barnaby:2011qe, Barnaby:2011vw, Anber:2012du, Linde:2012bt}, gravitational waves~\cite{Cook:2011hg, Barnaby:2011qe, Anber:2012du, Domcke:2016bkh, Bartolo:2016ami, Jimenez:2017cdr}, primordial black holes~\cite{Linde:2012bt, Bugaev:2013fya, Garcia-Bellido:2016dkw, McDonough:2016xvu, Domcke:2017fix, Garcia-Bellido:2017aan, Cheng:2018yyr}, $\mu$-distortions~\cite{Meerburg:2012id, Domcke:2016bkh}, primordial magnetic fields~\cite{Anber:2006xt, Durrer:2010mq, Caprini:2014mja, Fujita:2015iga, Green:2015fss, Patel:2019isj} and the generation of the baryon asymmetry~\cite{Anber:2015yca, Cado:2016kdp, Jimenez:2017cdr, Domcke:2018eki, Domcke:2019mnd}.

In order to explore the efficiency of preheating in models described by the action in \cref{eqn:action}, we consider a range of single-field inflationary potentials forming a representative sample of those considered by Planck~\cite{Ade:2015lrj,Akrami:2018odb}.
We explicitly study five models:\footnote{Although these models are not all pseudoscalar inflationary scenarios, in this Letter we are interested in the dependence of preheating on the potential shape, and thus ignore the detailed origin of the potentials.} \emph{chaotic inflation}~\cite{Linde:1983gd},
\begin{align}
    \label{eq:chaotic-class}
    V(\phi) = \frac{1}{2} m_\phi^2 \phi^2,
\end{align}
\emph{Starobinsky-like} models~\cite{Starobinsky:1980te},
\begin{align}
    \label{eq:starobinsky_class}
    V(\phi) = V_0 \left[ 1 - \exp\left( \frac{\vert \phi \vert}{v} \right) \right]^2,
\end{align}
the \emph{axion-monodromy} model~\cite{Silverstein:2008sg, McAllister:2008hb, Flauger:2009ab},
\begin{align}
    V(\phi) = \mu^3 \left( \sqrt{\phi^2 + \phi_c^2} - \phi_c \right),
\end{align}
\emph{hilltop-like} models~\cite{Boubekeur:2005zm},
\begin{align}
    V(\phi) = V_0 \left[ 1 - \left( \frac{\vert \phi \vert}{v} \right)^p \right]^2,
\end{align}
and \emph{D-brane} models~\cite{Dvali:2001fw, Burgess:2001fx, GarciaBellido:2001ky, Kachru:2003sx},
\begin{align}
    V(\phi) = V_0 \left[ 1 - \left( \frac{v}{\vert \phi \vert} \right)^p \right]^2.
\end{align}
We also consider natural inflation~\cite{Freese:1990rb}, $V(\phi) = V_0 \left[ 1 + \cos\left( \phi/v\right) \right]$,
but the results (for $v = \sqrt{8 \pi} \Mpl$) are virtually identical to those for chaotic inflation, so we omit them below.
In \cref{tab:model-params} we enumerate various model and simulation parameters and the predictions for inflationary observables.
In all cases, after fixing the free parameters denoted in \cref{tab:model-params}, the normalization of the scalar power spectrum~\cite{Akrami:2018odb} was used to fix the parameter determining the scale of the potential.
\begin{table*}[th]
    \centering
    \caption{The specific parameters chosen for each inflationary model under consideration.
        We report the effective inflaton mass, the simulation box length, the number of $e$-folds before the end of inflation we start the simulation, the Hubble rate at the end of inflation $H_e$, the ratio of the lattice's infrared cutoff to the comoving Hubble scale at the end of inflation, equal to $(2 \pi / L) / \mathcal{H}_e$, and the energy scale at the end of inflation.
        In addition, we list the tilt of the scalar power spectrum $n_s$ and the tensor-to-scalar ratio $r$, evaluated at a pivot scale which left the horizon 60 $e$-folds before inflation ended.
        }\label{tab:model-params}
    \begin{tabular*}{\textwidth}[t]{l @{\extracolsep{\fill}} cccccccc}
        \toprule
        Model &             $m_\phi / \Mpl$ &           $L m_\phi$ &    $N_0$ &
        $H_e / m_\phi$ & $k_\mathrm{IR} / \mathcal{H}_e$ &  $\sqrt[4]{\rho_e} / \Mpl$ &
        $n_s$     & $r$
        \\ \midrule
        Chaotic ($n=2$) & $6.16 \times 10^{-6}$ & 15 & $-2$ & 0.51 & 0.82 & $2.3 \times 10^{-3}$ & 0.966 & 0.13
        \\
        Starobinsky ($v = 10 \Mpl/3$) & $1.06 \times 10^{-5}$ & 20 & $-2$ & 0.37 & 0.85 & $2.6 \times 10^{-3}$ & 0.969 & 0.016
        \\
        Monodromy ($\phi_c = \Mpl /10$) & $4.66 \times 10^{-5}$ & 50 & $-2$ & 0.15 & 0.84 & $3.5 \times 10^{-3}$ & 0.975 & 0.067
        \\
        Hilltop ($p = 4$, $v = 4 \Mpl$) & $3.06 \times 10^{-6}$ & 20 & $-2$ & 0.24 & 1.3 & $1.1 \times 10^{-3}$ & 0.951 & $1.4 \times 10^{-4}$
        \\
        Hilltop ($p = 4$, $v = 2 \Mpl$) & $1.60 \times 10^{-6}$ & 20 & $-1$ & 0.15 & 2.1 & $6.5 \times 10^{-4}$ & 0.949 & $9.8 \times 10^{-6}$
        \\
        D-brane $(p=2, v=\Mpl/2)$ & $4.90 \times 10^{-5}$ & 40 & $-1$ & 0.073 & 2.1 & $2.5 \times 10^{-3}$ & 0.975 & $2.2 \times 10^{-3}$
        \\
        \bottomrule
    \end{tabular*}
\end{table*}

Gravitational waves correspond to the tensor component $h_{ij}$ of a general perturbation to a homogeneous spacetime,
\begin{align}
    \label{eq:spacetime_metric}
    \ud s^2
    &= a(\tau)^2 \left[ - \ud \tau^2 + \left(\delta_{ij} + h_{ij}\right) \ud x^i \ud x^j \right],
\end{align}
for which the linearized Einstein equation yields a second-order differential equation sourced by the transverse-traceless projection of the anisotropic stress tensor (see, e.g., Ref.~\cite{Adshead:2018doq}).
We evolve these tensor degrees of freedom in tandem with the axion and gauge field, extracting the spectrum of fractional energy density in gravitational waves,
\begin{align}\label{eqn:gw-spectrum}
	\Omega_\mathrm{gw}(k)
	&\equiv \frac{1}{\rho} \frac{\ud \rho_\mathrm{gw}}{\ud \ln k} \\
	&= \frac{1}{24\pi^2 L^3} \frac{k^3}{\mathcal{H}^2} \sum_{i, j} \left\vert \partial_0 h_{ij}(k, \tau) \right\vert^2,
\end{align}
where $L^3$ is the (comoving) simulation volume.
Integrating \cref{eqn:gw-spectrum} yields the net fraction of energy residing in gravitational waves, $\Omega_{\mathrm{gw}}$.
The bounds on $N_\mathrm{eff}$ place an upper bound on the energy density in radiation beyond the standard model, $\Delta N_\mathrm{eff} = N_\mathrm{eff} - 3.046$, which directly constrains the fraction of energy in gravitational waves today, $\Omega_{\mathrm{gw},0} h^2$, via~\cite{Maggiore:1999vm}
\begin{align}
	\frac{ \Omega_{\mathrm{gw},0} h^2}{\Omega_{\gamma,0} h^2} &= \frac{7}{8} \left( \frac{4}{11} \right)^{4/3} \Delta N_\mathrm{eff}.
\end{align}
In what follows, we compare the resulting bounds to the gravitational wave production from preheating.

%%%%%%%%%%%%%%%%%%%%%%%%%%%%%%%%%%%%%%%%%%%%%%%%%%%%%%%%%%%%%%%%%%%%%%%%%%%%%%%%%%%%

\textit{Results}.---Similarly to Ref.~\cite{Adshead:2018doq}, we numerically evolve the classical equations of motion of the gauge fields, \cref{eq:A_i-eom}, and axion, \cref{eq:phi-eom}, in an FLRW background.
The evolution equations are discretized onto a 3D, periodic, regularly spaced grid, using fourth-order centered differencing for spatial derivatives and the fourth-order Runge-Kutta method for time integration.
For this work we developed \textsf{pystella},\footnote{\href{https://github.com/zachjweiner/pystella}{github.com/zachjweiner/pystella}} an MPI-parallel and GPU-accelerated Python package which relies on \textsf{PyOpenCL}~\cite{kloeckner_pycuda_2012} and \textsf{Loo.py}~\cite{kloeckner_loopy_2014} for the generation and execution of OpenCL code on GPUs.
As such, \textsf{pystella} allows for reliable simulations of larger couplings $\alpha / f$ than in Ref.~\cite{Adshead:2018doq} using higher-resolution grids with $384^3$ points and a time step of $\Delta \tau = \Delta x / 10$~\cite{Adshead:2019lbr}.
For details on our procedure for setting initial conditions, refer to Appendix B of Ref.~\cite{Adshead:2019lbr}.

Changing the shape of a scalar field's potential changes its effective mass $m_\phi$ (defined by $m_\phi^2 = \partial^2 V / \partial \phi^2$ evaluated at the minimum of the potential), which sets the oscillation timescale for the axion background and determines the wave numbers of importance during preheating.
In particular, the ratio of the Hubble rate at the end of inflation to the axion's effective mass differs from model to model, requiring different comoving box sizes $L$ for sufficient long-wavelength resolution (listed in \cref{tab:model-params}).

\begin{figure}[t]
    \centering
    \includegraphics[width=\columnwidth]{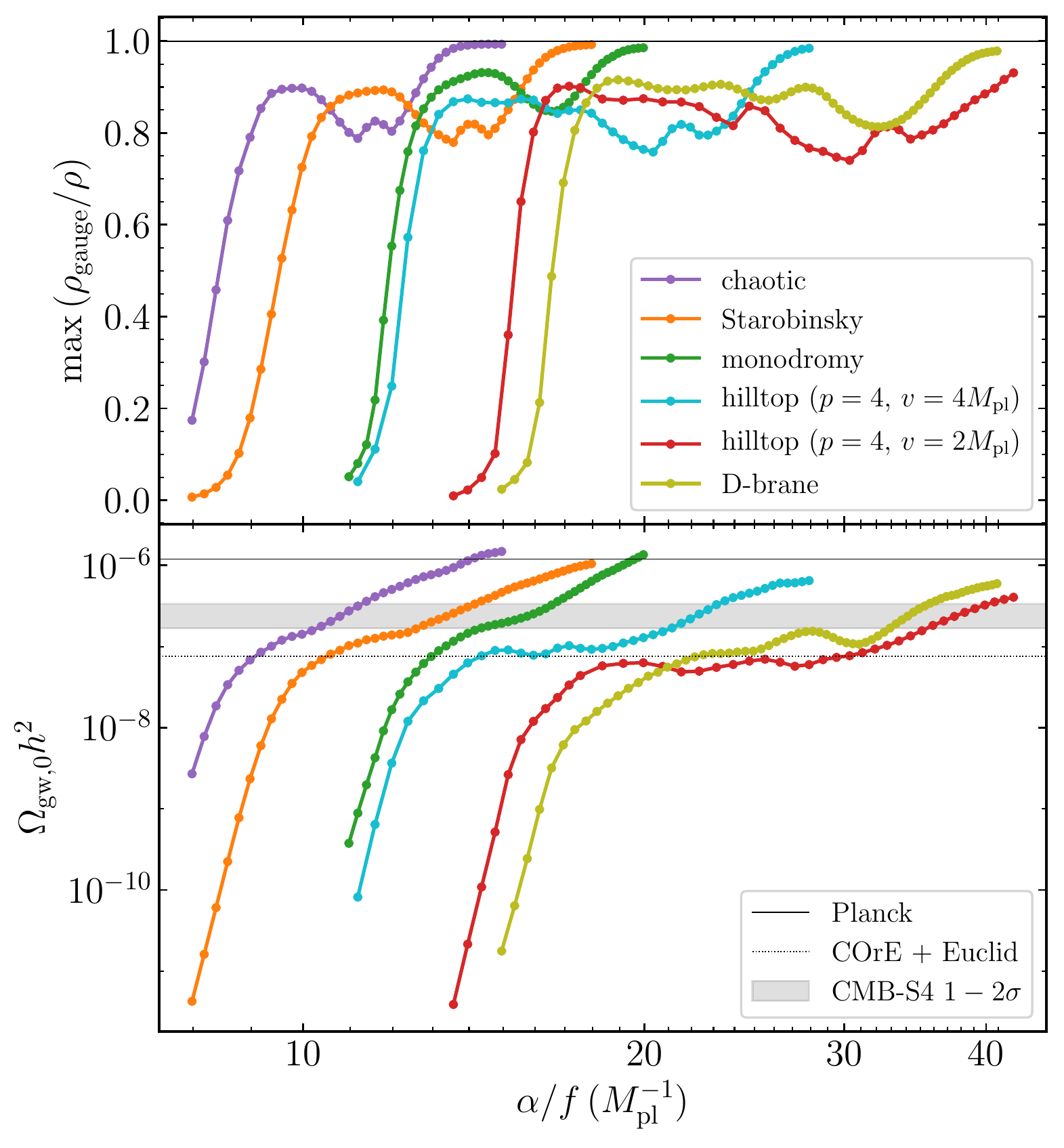}
    \caption{
        Preheating efficiency, quantified by the maximum $\rho_\mathrm{gauge} / \rho$ over the simulation (top panel), and the total fractional energy in gravitational waves today, $\Omega_{\mathrm{gw}, 0} h^2$ (bottom panel), as functions of axion-gauge coupling $\alpha / f$.
        Lines indicating $\Delta N_\mathrm{eff}$ bounds on $\Omega_{\mathrm{gw}, 0} h^2$ from Planck and CMB-S4 from Ref.~\cite{Pagano:2015hma} are plotted in solid and dashed black, respectively, while the region between CMB-S4's $1 \sigma$ and $2 \sigma$ projections~\cite{Abazajian:2019eic} is shaded grey.
    }\label{fig:gw-money-vs-model}
\end{figure}

In \cref{fig:gw-money-vs-model} we study the relationship between gravitational wave production and the efficiency of preheating, quantified by the maximum fraction of energy in the gauge fields over the simulation.
The top panel shows that the relationship between preheating efficiency and the coupling $\alpha / f$ follows a similar trend regardless of the inflationary potential (though this trend manifests at different values of $\alpha / f$ for different models).
The bottom panel of \cref{fig:gw-money-vs-model} shows that (at sufficiently high coupling) preheating in all models produces gravitational waves that would be probed by CMB-S4, while models with tensor-to-scalar ratios $r \gtrsim 10^{-2}$ are already limited by Planck data~\cite{Pagano:2015hma}.

While all models exhibit a similar relationship between preheating efficiency and gravitational wave production, some models result in larger overall $\Omega_{\mathrm{gw}, 0} h^2$.
This difference is due in part to the differing location of the  peak of the gravitational wave source relative to the horizon. Because lower-scale inflationary models require larger couplings $\alpha / f$ for preheating to be comparably efficient to high-scale models, gauge-field modes deeper within the horizon are more strongly amplified relative to those in higher-scale models~\cite{Adshead:2019lbr}.
Following a ``rule of thumb'' for cosmological stochastic gravitational wave backgrounds~\cite{Giblin:2014gra}, the peak amplitude of a gravitational wave signal is suppressed if its source is further inside the horizon.
Consulting \cref{tab:model-params}, we observe that models with large tensor-to-scalar ratios ($r \gtrsim 10^{-2}$) preheat efficiently at lower coupling, and subsequently exhibit higher levels of gravitational wave production.
Since $r$ measures the energy scale of inflation, models with smaller $r$ require larger coupling for complete preheating, resulting in smaller $\Omega_{\mathrm{gw}}$ even if preheating itself is equally efficient.

These results demonstrate that for inflationary potentials whose tensor-to-scalar ratios would be observable by CMB-S4 experiments, the entire regime of efficient gauge preheating ($\gtrsim 80$\% efficiency) will be probed via the contribution of gravitational waves to $\Delta N_\mathrm{eff}$.
On the one hand, a detection of both $r$ and $\Delta N_\mathrm{eff}$ would be consistent with a pseudoscalar inflaton strongly preheating to gauge fields; on the other hand, nearly the entire regime of efficient preheating would be ruled out by a null measurement of $\Delta N_\mathrm{eff}$, leading to upper bounds on the axion-gauge coupling $\alpha / f$ in all models.
Similarly, in the event that next generation experiments limit $r < 10^{-3}$, a detection of nonzero $\Delta N_\mathrm{eff}$ is consistent with efficient gauge preheating.

\begin{figure}[t]
    \centering
    \includegraphics[width=\columnwidth]{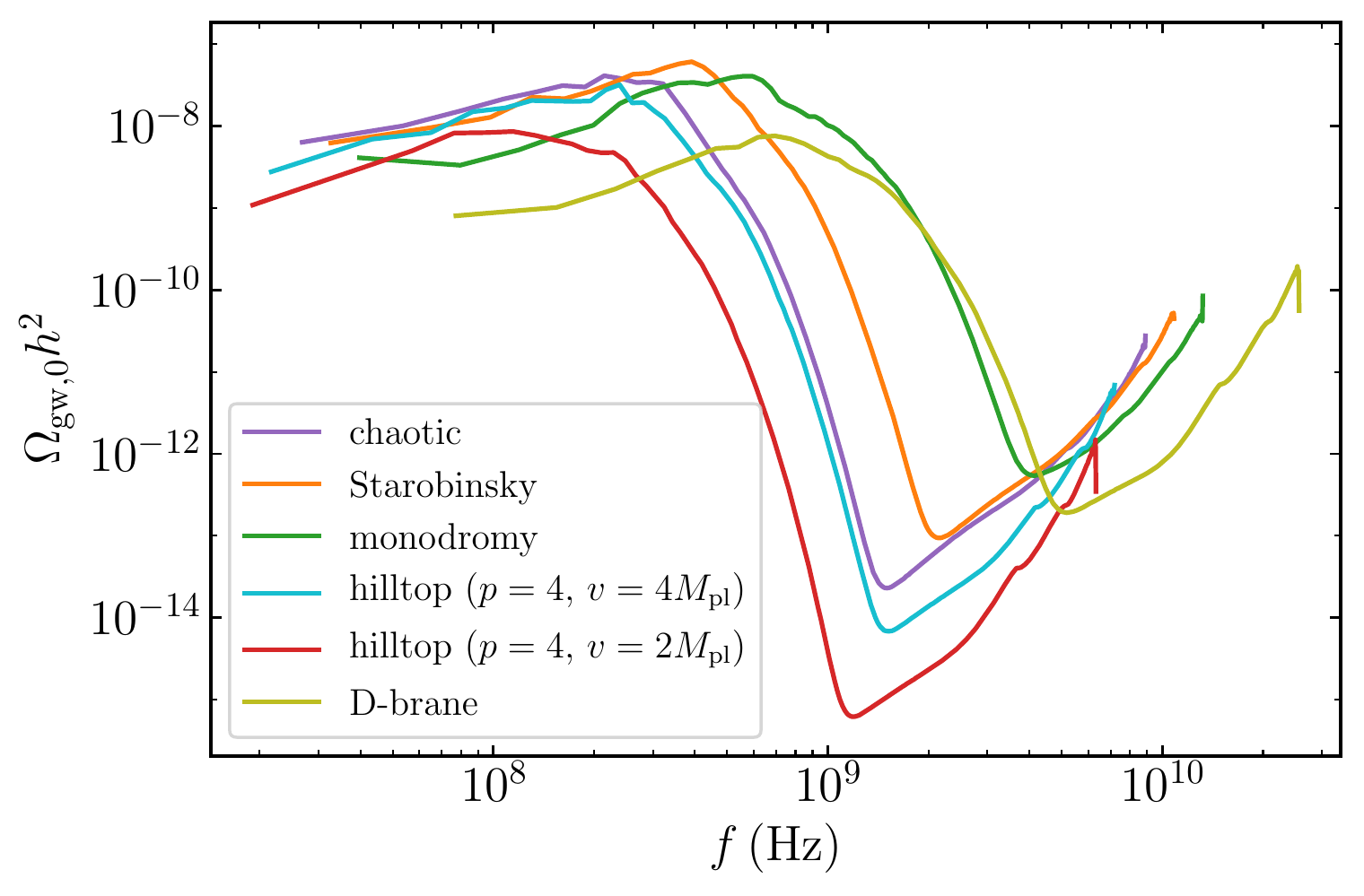}
    \caption{
        The present-day gravitational wave spectra resulting from gauge preheating after inflation with each potential listed in \cref{tab:model-params} (with colors denoted by the legend), plotted against the frequencies which would be observed today.
        The coupling in each case is the smallest value simulated for which (a maximum of) $85\%$ of the energy in the simulation ends up in the gauge fields.
        Note that the ultraviolet parts of the spectra growing with $k^4$ result from vacuum modes in the simulation and so are not physical signals.
    }\label{fig:gw-spectra-vs-f}
\end{figure}

In \cref{fig:gw-spectra-vs-f}, we plot the gravitational wave spectra that would be observed today as a stochastic background, where the amplitude of the signal at emission $\Omega_{\mathrm{gw},e}(f)$ would have redshifted to $\Omega_\mathrm{gw,0}(f) h^2 = \Omega_{\mathrm{gw},e}(f) \left( g_0 / g_\ast \right)^{1/3} \Omega_{\mathrm{r},0} h^2$~\cite{Easther:2006vd}.
The shapes of the signals from preheating in all inflationary scenarios are broadly similar, exhibiting the single broad peak characteristic to tachyonic resonances, though the frequency of this peak varies from model to model.
The present-day frequency of emission corresponding to a physical wave number $k_\mathrm{phys}$ is $f = 2.7 \times 10^{10} k_\mathrm{phys} / \sqrt{\Mpl H} \, \mathrm{Hz}$, where $H$ is the Hubble parameter at the time of emission~\cite{Easther:2006gt}.
Because the wave numbers important for preheating are $k \sim m_\phi$, the relevant frequencies for a given (inflationary) model scale with $m_\phi / \sqrt[4]{\rho}$, which is reflected by the peak locations in \cref{fig:gw-spectra-vs-f}.
The signals in \cref{fig:gw-spectra-vs-f} peak at $\Omega_{\mathrm{gw}, 0} \sim 10^{-7}$, while for the most efficient couplings studied here the signals approach $10^{-6}$.

%%%%%%%%%%%%%%%%%%%%%%%%%%%%%%%%%%%%%%%%%%%%%%%%%%%%%%%%%%%%%%%%%%%%%%%%%%%%%%%%%%%%

\textit{Conclusions}.---A dramatic stochastic background of gravitational waves is generated by the resonant amplification of Abelian gauge fields coupled to a pseudoscalar inflaton (axion).
The net radiation in gravitational waves, propagated to the present day, is so great that it provides the strongest constraints on the axion-gauge coupling $\alpha / f$.
While the quantitative constraints on $\alpha / f$ depend on the inflationary potential, a measurement of $\Delta N_\mathrm{eff}$ consistent with zero by next-generation experiments would all but rule out the regime in which the Universe was reheated by gauge preheating alone in high-scale inflation.
In this Letter we have demonstrated that this result is qualitatively generic across (a representative sample of) single-field models of inflation, highlighting that the greatest detection prospects coincide with models whose tensor-to-scalar ratio would also be detected by CMB-S4.
Combining constraints on the tensor-to-scalar ratio with constraints from the gravitational wave contribution to $N_\mathrm{eff}$ constrains models of axion inflation across the most disparate scales available, spanning 29 decades in frequency.

This result represents the first observational constraints from preheating.
In particular, the constraints on the inflaton--gauge-field coupling provided by gravitational waves from preheating are tighter than those from primordial black hole production~\cite{Linde:2012bt, Bugaev:2013fya}, which constrain $\alpha / f \lesssim 21.9 \Mpl^{-1} - 24.9 \, \Mpl^{-1}$ and $\alpha / f \lesssim 35.9 \, \Mpl^{-1}$ for the chaotic and monodromy potentials, respectively.
(The corresponding constraints from non-Gaussianity~\cite{Barnaby:2010vf,Barnaby:2011vw} are $\alpha / f \lesssim 32.3 \Mpl^{-1}$ and $\alpha / f \lesssim 46.5 \, \Mpl^{-1}$.)
Our results limit $\alpha / f \lesssim 14 \, \Mpl^{-1}$ and $19.6 \, \Mpl^{-1}$ for these two potentials, while next-generation experiments could limit $\alpha / f \lesssim 9 \, \Mpl^{-1}$ and $13 \, \Mpl^{-1}$, respectively.
These results also have implications for constraining models of dark photon dark matter~\cite{Bastero-Gil:2018uel,Agrawal:2018vin,Co:2018lka,Dror:2018pdh,Machado:2018nqk}.

Our findings suggest that gauge preheating may result in strong, nonlinear gravitational effects, prompting future study into gravitational backreaction from metric perturbations, or even using Numerical Relativity as recently employed for scalar-field preheating~\cite{Giblin:2019nuv}.
At couplings even stronger than considered here, the friction the gauge-fields exert on the axion background may delay the end of inflation, which could amplify the production of primordial black holes.
We defer these investigations to future work.

%%%%%%%%%%%%%%%%%%%%%%%%%%%%%%%%%%%%%%%%%%%%%%%%%%%%%%%%%%%%%%%%%%%%%%%%%%%%%%%%%%%%

We thank Mustafa Amin and Valerie Domcke for useful discussions and comments on the draft.
Z.J.W. thanks Andreas Kloeckner for generous support and advice on the development of \textsf{pystella}.
The work of P.A.\ was supported in part by NASA Astrophysics Theory Grant No. NNX17AG48G.
J.T.G.\ is supported by the National Science Foundation Grant No. PHY-1719652.
M.P.\ acknowledges the support of the Spanish MINECOs ``Centro de Excelencia Severo Ochoa'' Programme under Grant No. SEV-2016-059.
This project has received funding from the European Unions Horizon 2020 research and innovation programme under the Marie Sk\l{}odowska-Curie Grant No. 713366.
Z.J.W.\ is supported in part by the United States Department of Energy Computational Science Graduate Fellowship, provided under Grant No. DE-FG02-97ER25308.
The development of \textsf{pystella} made use of the Extreme Science and Engineering Discovery Environment (XSEDE)~\cite{xsede} through allocation TG-PHY180049, which is supported by National Science Foundation Grant No. ACI-1548562, and also made use of hardware purchased by the National Science Foundation, Kenyon College, and the Kenyon College Department of Physics.
This work made use of the Illinois Campus Cluster, a computing resource that is operated by the Illinois Campus Cluster Program (ICCP) in conjunction with the National Center for Supercomputing Applications (NCSA) and which is supported by funds from the University of Illinois at Urbana-Champaign.
P.A.\ acknowledges the hospitality of the Yukawa Institute for Theoretical Physics at Kyoto University, where some of this work was completed during the YITP-T-19-02 on ``Resonant instabilities in cosmology.''

\bibliography{axion-gw}

\end{document}